\input harvmac
\input epsf

\newcount\figno
\figno=0 
\def\fig#1#2#3{
\par\begingroup\parindent=0pt\leftskip=1cm\rightskip=1cm\parindent=0pt
\baselineskip=11pt
\global\advance\figno by 1
\midinsert
\epsfxsize=#3
\centerline{\epsfbox{#2}}
\vskip 12pt
{\bf Fig.\ \the\figno: } #1\par
\endinsert\endgroup\par
}
\def\figlabel#1{\xdef#1{\the\figno}}

\skip0=\baselineskip
\divide\skip0 by 2

\def\IR{\relax{\rm I\kern-.18em R}}
\def\IZ{\relax\ifmmode\hbox{Z\kern-.4em Z}\else{Z\kern-.4em Z}\fi}
\def\IQ{\relax{\rm I\kern-.40em Q}}
\def\IS{\relax{\rm I\kern-.18em S}}
\def\p{\partial}

\def\msurr{\mathsurround=0pt}
\def\overleftrightarrow#1{\vbox{\msurr\ialign{##\crcr
        $\leftrightarrow$\crcr\noalign{\kern-1pt\nointerlineskip}
        $\hfil\displaystyle{#1}\hfil$\crcr}}}

\def\frac#1#2{{#1\over #2}}
\def\a{{\alpha}}

\lref\btz{
M.~Banados, C.~Teitelboim and J.~Zanelli,
``The Black hole in three-dimensional space-time,''
Phys.\ Rev.\ Lett.\  {\bf 69}, 1849 (1992).
}
\lref\banados{
M.~Banados, M.~Henneaux, C.~Teitelboim and J.~Zanelli,
``Geometry of the (2+1) black hole,''
Phys.\ Rev.\ D {\bf 48}, 1506 (1993).
}
\lref\bala{
V.~Balasubramanian, J.~de Boer, E.~Keski-Vakkuri and S.~F.~Ross,
``Supersymmetric conical defects: Towards a string theoretic description  of black hole formation,''
hep-th/0011217.
}
\lref\horo{
G.~T.~Horowitz and D.~L.~Welch,
``String Theory Formulation of the Three-Dimensional Black Hole,''
Phys.\ Rev.\ Lett.\  {\bf 71}, 328 (1993).
}
\lref\malstr{
J.~Maldacena and A.~Strominger,
``AdS(3) black holes and a stringy exclusion principle,''
JHEP {\bf 9812}, 005 (1998).
}
\lref\bh{
J.~D.~Brown and M.~Henneaux,
``Central Charges In The Canonical Realization Of Asymptotic Symmetries: An Example From Three-Dimensional Gravity,''
Commun.\ Math.\ Phys.\  {\bf 104}, 207 (1986).
}
\lref\as{A.~Strominger,
``Black hole entropy from near-horizon microstates,''
JHEP {\bf 9802}, 009 (1998).
}
\lref\adscft{
O.~Aharony, S.~S.~Gubser, J.~Maldacena, H.~Ooguri and Y.~Oz,
``Large N field theories, string theory and gravity,''
Phys.\ Rept.\  {\bf 323}, 183 (2000).
}
\lref\vafwit{
C.~Vafa and E.~Witten,
``Bosonic String Algebras,''
Phys.\ Lett.\ B {\bf 159}, 265 (1985).
}
\lref\govin{
T.~R.~Govindarajan, T.~Jayaraman, A.~Mukherjee and S.~R.~Wadia,
``Twisted Current Algebras And Gauge Symmetry Breaking In String Theory,''
Mod.\ Phys.\ Lett.\ A {\bf 1}, 29 (1986).
}
\lref\mo{
J.~Maldacena and H.~Ooguri,
``Strings in AdS(3) and the SL(2,R) WZW model. I:  The spectrum,''
hep-th/0001053.
}
\lref\mos{
J.~Maldacena, H.~Ooguri and J.~Son,
``Strings in AdS(3) and the SL(2,R) WZW model. II: Euclidean black hole,''
hep-th/0005183.
}
\lref\polchin{
J.~Polchinski,
``Evaluation Of The One Loop String Path Integral,''
Commun.\ Math.\ Phys.\  {\bf 104}, 37 (1986).
}
\lref\mcclain{
B.~McClain and B.~D.~Roth,
``Modular Invariance For Interacting Bosonic Strings At Finite Temperature,''
Commun.\ Math.\ Phys.\  {\bf 111}, 539 (1987).
}
\lref\dlp{
L.~J.~Dixon, M.~E.~Peskin and J.~Lykken,
``N=2 Superconformal Symmetry And SO(2,1) Current Algebra,''
Nucl.\ Phys.\ B {\bf 325}, 329 (1989).
}
\lref\mms{
J.~Maldacena, J.~Michelson and A.~Strominger,
``Anti-de Sitter fragmentation,''
JHEP {\bf 9902}, 011 (1999).
}
\lref\seiwit{
N.~Seiberg and E.~Witten,
``The D1/D5 system and singular CFT,''
JHEP {\bf 9904}, 017 (1999).
}
\lref\mart{
E.~Martinec and W.~McElgin,
``String theory on AdS orbifolds,''
hep-th/0106171.
}
\lref\balog{
J.~Balog, L.~O'Raifeartaigh, P.~Forgacs and A.~Wipf,
``Consistency Of String Propagation On Curved Space-Times: An SU(1,1) Based Counterexample,''
Nucl.\ Phys.\ B {\bf 325}, 225 (1989).
}
\lref\petro{
P.~M.~Petropoulos,
``Comments On SU(1,1) String Theory,''
Phys.\ Lett.\ B {\bf 236}, 151 (1990).
}
\lref\moham{
N.~Mohammedi,
``On The Unitarity Of String Propagation On SU(1,1),''
Int.\ J.\ Mod.\ Phys.\ A {\bf 5}, 3201 (1990).
}
\lref\barsnem{
I.~Bars and D.~Nemeschansky,
``String Propagation In Backgrounds With Curved Space-Time,''
Nucl.\ Phys.\ B {\bf 348}, 89 (1991).
}
\lref\hwanga{
S.~Hwang,
``No ghost theorem for SU(1,1) string theories,''
Nucl.\ Phys.\ B {\bf 354}, 100 (1991).
}
\lref\hwangb{
S.~Hwang,
``Cosets as gauge slices in SU(1,1) strings,''
Phys.\ Lett.\ B {\bf 276}, 451 (1992).
}
\lref\hwangc{
S.~Hwang,
``Unitarity of strings and non-compact Hermitian symmetric spaces,''
Phys.\ Lett.\ B {\bf 435}, 331 (1998).
}
\lref\evans{
J.~M.~Evans, M.~R.~Gaberdiel and M.~J.~Perry,
``The no-ghost theorem for AdS(3) and the stringy exclusion principle,''
Nucl.\ Phys.\ B {\bf 535}, 152 (1998).
}
\lref\deser{
S.~Deser and R.~Jackiw,
``Three-Dimensional Cosmological Gravity: Dynamics Of Constant Curvature,''
Annals Phys.\  {\bf 153}, 405 (1984).
}
\lref\gawed{
K.~Gawedzki,
``Noncompact WZW conformal field theories,''
hep-th/9110076.
}
\lref\lyk{
J.~D.~Lykken,
``Finitely Reducible Realizations Of The N=2 Superconformal Algebra,''
Nucl.\ Phys.\ B {\bf 313}, 473 (1989).
}
\lref\argurio{
R.~Argurio, A.~Giveon and A.~Shomer,
``Superstrings on AdS(3) and symmetric products,''
JHEP {\bf 0012}, 003 (2000).
}
\lref\mezin{
L.~Mezincescu and P.~K.~Townsend,
``Stability At A Local Maximum In Higher Dimensional Anti-De Sitter Space And Applications To Supergravity,''
Annals Phys.\  {\bf 160}, 406 (1985).
}
\lref\gep{
D.~Gepner and Z.~Qiu,
``Modular Invariant Partition Functions For Parafermionic Field Theories,''
Nucl.\ Phys.\ B {\bf 285}, 423 (1987).
}

\Title{\vbox{\baselineskip12pt\hbox{hep-th/0107131}\hbox{HUTP-01/A034}
\hbox{}}}{String Theory on $AdS_3/Z_N$}

\centerline{John Son}
\bigskip\centerline{Department of Physics}
\centerline{Harvard University} 
\centerline{Cambridge, MA 02138}
\bigskip\centerline{{\tt json@pauli.harvard.edu}}

\vskip .3in \centerline{\bf Abstract}

We study string theory on singular $Z_N$ quotients of $AdS_3$, corresponding
to spaces with conical defects.  The spectrum is computed using the orbifold
procedure.  It is shown that spectral flow may be used to generate
the twisted sectors.  We further compute the thermal partition function and 
show that it correctly reproduces the spectrum. 

\smallskip 
\Date{}

\newsec{Introduction}

General Relativity in three dimensions  with a negative cosmological 
constant,
\eqn\graction{
S = {1 \over 2 \pi} \int d^3x \sqrt{-g} \left( R + {2 \over l^2} \right) +
{\rm surface \ terms} \;,
}
has a family of solutions labelled by two parameters $M$ and $J$ 
\refs{\btz, \banados}
\eqn\famsl{\eqalign{
ds^2 & = -N^2 dt^2  + N^{-2} dr^2 + r^2(N^{\phi} dt + d\phi)^2 \;, \cr
N^2 & =  -M + {r^2 \over l^2} + {J^2 \over 4r^2} \;, \cr
N^{\phi} & = -{J \over 2r^2}\;.
}}
What the resulting spacetimes look like depend on the values of the 
two parameters.
When $M > 0$ and $ Ml > |J|$, these spacetimes correspond to black holes.  
The second condition ensures that a horizon exists.  The constants are then
identified with the mass and angular momentum of the black hole, respectively.
These spaces may be thought of as excitations of the $M=0$ case.  

However, $M=0$ is not the lowest energy state possible.  It is known that
when $M=-1$ the space \famsl\ corresponds to the three-dimensional 
anti-de Sitter space, $AdS_3$.  

For the spacetimes with $-1 < M < 0$ (and $J=0$), a rescaling of the 
coordinates brings the metric into the form
\eqn\metric{
ds^2 = - \left(1 + {r^2 \over l^2}\right) dt^2 + 
\left(1 + {r^2 \over l^2}\right)^{-1} dr^2  + r^2 d\phi^2 \;,
}    
which is the same as $AdS_3$, but with a deficit angle $\delta = 2 \pi (1-\sqrt{|M|})$ for $\phi$.  Thus, these spaces correspond to $AdS_3$ with 
conical singularities.

In fact, even the black holes corresponding to \famsl\ with $M>0$ are locally
$AdS_3$, and can be obtained from $AdS_3$ by a quotient.  This is consistent
with the equations of motion resulting from \graction, which 
implies that the curvature is constant.  The black hole solutions do not 
have a curvature singularity, 
and differ from $AdS_3$ only by some global identifications. 

The solutions that are being discussed here are easily lifted to solutions
of string theory.  By including a three form $H$, which must be proportional 
to the volume form in three dimensions, these spaces provide a background
in which it is possible to describe string propagation via the $SL(2,R)$
WZW model.  At the level of low energy effective action, \graction\ arises
when one takes the action for the massless fields of string theory
$g_{\mu \nu}$, $H$, $\phi$ and sets  
$H_{\mu \nu \sigma} = {2 \over l} \epsilon_{\mu \nu \sigma}$, and
$\phi =0$ \refs{\horo}.

The purpose of this paper is to study string propagation on the conical spaces.
For the special values of the opening angle  ${2\pi/ N}$,
where $N$ is an integer, the spaces may be obtained as a $Z_N$ orbifold of 
$AdS_3$.\foot{$\phi$ corresponds to rotation in $X-Y$ plane in the 
covering space 
$ds^2 = - dU^2 - dV^2 + dX^2 + dY^2$,
and is always a space-like killing vector, ensuring causality in the 
resulting quotient space.}      
The singularity present is then just an orbifold singularity, and it is 
possible to formulate a consistent string theory on this background given the 
knowledge of string theory on $AdS_3$.

There are many reasons for studying this theory.  Such spaces can be 
formed by adding mass to empty $AdS_3$ \refs{\deser}.  Relative to the 
$AdS_3$ vacuum, an object of mass less than 1 would create a conical
singularity.  
One can imagine a process where a collision taking place
inside of $AdS_3$ leaves a lump of stable matter, not enough to produce
a black hole but distorting the geometry to what we are studying here.  Indeed,
this provides a controlled setting to study black hole formation, as in 
\refs{\bala}.

Another reason is that a consistent description of strings propagating
on $AdS_3$ has emerged only recently in \refs{\mo}.  The crucial observation
made in that work is the existence of spectral flow as a symmetry of the
$SL(2,R)$ WZW model.  One would like to gain further insight into this
spectral flow symmetry.  In this paper we will learn that on the conical
spaces spectral flow acts as a twist, in the orbifold sense.

Finally, $AdS_3$/CFT$_2$ correspondence \refs{\malstr, \bh, \as} stands 
apart from the correspondence in other dimensions \refs{\adscft}
in that it may be studied in a fully string-theoretic 
context, and not just in the gravity approximation.  
Therefore, it will be useful to understand as much as possible string theory 
on $AdS_3$ and its extensions.

After this work was completed, we received \refs{\mart} which has some overlap
with this paper.

\newsec{String theory on $AdS_3$} 

We begin by briefly summarizing the known results of string theory on 
$AdS_3$ \refs{\mo},
as this will serve as the theory on the covering space of the orbifold.
From this point on, we set $l=1$.

The action is given by that of the $SL(2,R)$ WZW model
\eqn\wzwac{
S =  {k \over 8 \pi} \int d^2 \sigma \ {\rm Tr} \ 
(g^{-1} \p _a g g^{-1} \p ^a g)
+{ik \over 12 \pi} \int {\rm Tr} \ (g^{-1}dg \wedge  g^{-1}dg \wedge  g^{-1}dg)
}
where the level $k$ need not be quantized as $H^3$ vanishes for $SL(2,R)$.
We choose the parametrization for the group element
\eqn\grp{
g = e^{iu \sigma_2} e^{\rho \sigma_3} e^{iv\sigma_2},
}  
where $\sigma$'s are the Pauli matrices, and upon 
setting $u=\frac12(t+\phi), v=\frac12(t-\phi)$ we obtain 
the coordinates of \metric\ by the transformation
$r = \sinh \rho$.  To avoid closed timelike curves 
we take the universal cover of this space to be $AdS_3$.

The key property of the WZW model is invariance under the action of 
left and right multiplication by group elements.  This implies the existence 
of two sets of conserved currents, the right-moving
\eqn\concur{\eqalign{
J^3_R &= k(\p _+u + \cosh 2 \rho \ \p_+v) \;, \cr 
J^{\pm}_R & = k(\p _+ \rho \pm i \sinh 2 \rho \ \p_+ v)e^{\mp i2u} \;,
}}
and the left-moving counter parts which are obtained by the replacement
$\p_+ \to \p_-$ and $u \leftrightarrow v$.  The zero modes of $J^3_R$ and
$J^3_L$ will be denoted $J^3_0$ and $\bar{J}^3_0$ respectively, and they are
related to the spacetime energy and angular momentum of a state in $AdS_3$ by
\eqn\spel{\eqalign{
J^3_0 & = \frac12 (E + \ell) \;, \cr 
\bar{J}^3_0 & = \frac12 (E - \ell) \;.
}}
The modes of $J_{R, L}$ each generate the current algebra 
$\widehat{SL} (2, R)$.  Hence, the Hilbert space of $SL(2, R)$ WZW 
model is a sum of representations of 
$\widehat{SL} (2, R)_L \times \widehat{SL} (2, R)_R$.  The Hilbert
space of string theory on $AdS_3$ is the subspace obtained after imposing
 the Virasoro
constraints $(L_n - \delta_{n,0})|{\rm physical}\rangle = 0$.  In general,
we will consider a spacetime of the form $AdS_3 \times {\cal M}$ in which
${\cal M}$ is described by an appropriate CFT.

The representations that appear in the Hilbert space of $SL(2,R)$ WZW 
model are the following.  Start with the discrete representations 
$\hat{{\cal D}}^{+}_j \times \hat{{\cal D}}^{+}_j$, where 
$\hat{{\cal D}}^{+}_j$ is the representation obtained by acting with the 
raising operators $J^{3, \pm}_{n<0}$ on the unitary $SL(2, R)$ 
representation $D^+_j$, \foot{See \refs{\dlp} for details on 
representations of $SL(2, R)$.}  
and also include the representations obtained by 
spectral flow (same amount on left and right) \refs{\mo}
\eqn\wgen{\eqalign{ 
J^3_n &\rightarrow  \tilde{J}^3_n = J^3_n - {k \over 2}w \delta _{n,0}  \cr
J^+_n &\rightarrow  \tilde{J}^+_n = J^+_{n+w}   \cr
J^-_n &\rightarrow  \tilde{J}^-_n = J^-_{n-w}\;,
}}
which changes the Virasoro generators to 
\eqn\wvir{
\tilde{L}_n = L_n + w J^3_n -{k \over 4} w^2 \delta_{n,0} \;.
}
The resulting representations are denoted 
$\hat{{\cal D}}^{+,w}_j \times \hat{{\cal D}}^{+,w}_j$.  Spectral flow
by $-1$ gives the charge conjugated representations, 
$\hat{{\cal D}}^{+,w=-1}_j = \hat{{\cal D}}^{-}_{k/2-j}$.  In all these 
representations the $SL(2,R)$
spin $j$ must be in the range $\frac12 < j < {k-1 \over 2}$,
which is more restrictive than what is allowed by the no-ghost theorem 
\refs{\balog, \petro, \moham, \barsnem, \hwanga, \hwangb, \hwangc, \evans,
\dlp}.
In the context of string theory on $AdS_3$, these representations correspond 
to the short strings.  A state is labelled as 
$|j, m, \bar{m}, N, \bar{N}, h, \bar{h}\rangle$, where $m$ is the $J^3_0$ eigenvalue given by $j+q$ where
$q$ is some integer.  $N$ denotes the level of the current algebra and $h$
is the conformal weight coming from the CFT of ${\cal M}$.  Imposing the 
Virasoro constraints and using \spel\ the energy of this state is found to
be\foot{This equation does not look symmetric between the 
left and right quantum numbers because the level matching condition implies
that the expression in the square root is the same for left and right.}
\eqn\shorte{
E = 1 + q + \bar{q} + 2w + 
\sqrt{1 + 4(k-2)\left(N -wq + h -1 - \frac12 w(w+1)\right)} \;.
}
The other representations in the Hilbert space of $SL(2, R)$ WZW model 
are the spectral flows \wgen\ of continuous representations 
$\hat{{\cal C}}^{\a,w}_{1/2 +is} \times \hat{{\cal C}}^{\a,w}_{1/2 +is}$
where 
$\hat{{\cal C}}^{\a}_{1/2 +is}$ is built with $J^{3, \pm}_{n<0}$ acting on
the unitary  $SL(2, R)$ representation ${\cal C}^{\a}_{j=1/2+is}$.  
These states correspond to the long strings in $AdS_3$ \refs{\mms, \seiwit}, 
and $s$ is a real number representing the momentum in the radial direction. 
For the continuous representations, there is no relation between
$m$ and $j$.  Therefore the expression for energy $J^3_0 + \bar{J}^3_0$ 
will look different.  It is given by
\eqn\longe{
E = {k \over 2} w + {1 \over w} 
\left( {2s^2+\frac12 \over k-2 } + N +h +\bar{N}+\bar{h} -2 \right) \;.
}

It is possible to give an independent derivation of this spectrum, via 
a path integral on thermal $AdS_3$ \refs{\mos}.  
Going to Euclidean coordinates in both the worldsheet and spacetime, one 
periodically identifies time with period $\beta= 1/T$.  
Then one evaluates the 1-loop path integral including the topologically
non-trivial sectors corresponding to a string wrapping the thermal circle.
After integrating over the torus moduli, the amplitude is identified 
with the spacetime free energy of string states in Lorentzian $AdS_3$.
In this way one can simply read off the spectrum which agrees with
\shorte\ and \longe .

\newsec{Conical spaces as orbifolds of $AdS_3$}

\subsec{$Z_N$ quotient}

Taking string theory on $AdS_3$ as the starting point, the conical spaces
with opening angles $2 \pi / N$ are obtained by taking a $Z_N$ orbifold.  
Let us first note how spectral flow acts on this quotient space.  
The effect of spectral flow is to take a solution of the 
WZW equation of motion
\eqn\solg{
g = g_+(x^+)g_-(x^-)
}
and generate a new solution \refs{\mo}
\eqn\solw{
g_+(x^+) \to e^{{i \over 2}wx^+ \sigma_2} g_+(x^+) \;, \ \ \ \ 
g_-(x^-) \to g_-(x^-)e^{{i \over 2}wx^- \sigma_2} \;.
}
Under this operation, $t \to t + w \tau$ and $\phi \to \phi + w \sigma$.  
In regular $AdS_3$ closure of the string worldsheet required that $w$ 
be an integer, but now we see that $w$ only needs to be a multiple of $1/N$.

When we spectral flow by a fractional amount the $\widehat{SL}(2, R)$ currents
obey twisted boundary conditions. Consider the $n$th twisted sector:
\eqn\nsector{
J^+(x^+ +2\pi) = J^+(x^+) \ e^{-2\pi i n /N} \;, \ \ \ 
J^-(x^+ +2\pi) = J^-(x^+) \ e^{2\pi i n /N} \;.
}
Then the mode expansion is
\eqn\expan{
J^+ (z)= \sum_{r \in Z + n/N} J^+_r z^{-r-1} \;, \ \ \ \ \
J^- (z)= \sum_{s \in Z - n/N} J^-_s z^{-s-1} \;,
}
where $z=e^{ix^+}$.  The commutation relations are
\eqn\comm{\eqalign{
\left[ J^+_r, J^-_s \right] & =  -2J^3_{r+s} + kr \delta_{r+s} \cr
\left[ J^3_m, J^{\pm}_r \right] & =  \pm J^{\pm}_{m+r}   \cr
\left[ J^3_m, J^3_l \right] & =  -{k \over 2}m \delta_{m+l}\;. 
}}
There is a total of  $N$ sectors to consider, and in each sector $J^{\pm}$ are
quantized with different periodicity.  We now turn to the first step in 
taking an orbifold, which is to construct the twisted states.  Once again,
spectral flow will be seen to play a crucial role.

\subsec{Twisted states and spectral flow}

Consider a state obtained by repeated applications of the raising operators on
a lowest weight state, 
\eqn\tstate{
\prod_{m_i} J^3_{m_i}  \prod_{r_j} J^+_{r_j} 
\prod_{s_k} J^-_{s_k} |j,j\rangle \;.
}
If necessary, commutation relations may be used to change the order in which
the generators appear.  However, in what follows the ordering will be 
immaterial.  Such a state has $L_0 = -(\sum m_i + \sum r_j + \sum s_k)$ and
$J^3_0 = j + N^+ - N^-$ where $N^+$ ($N^-$) is the number of times 
$J^+$ ($J^-$) appears in the above expression.  Also note that the fractional
part of the level is given by $(N^+ - N^-)n/N$.
If we take this state and spectral flow \wgen\ by $w=-n/N$, we
find that the new generators acting on it are integrally moded.  
Thus, one can think of
this state as belonging to $\hat{{\cal D}}^{+,w=n/N}_{\tilde{j}}$.  
To obtain a string state in 
spacetime (including ${\cal M}$), we impose the Virasoro constraints 
\eqn\physt{\eqalign{
(L_0 -1)|\tilde{j}, \tilde{m}, \tilde{N}, h\rangle  & =
\left( -{\tilde{j}(\tilde{j} -1) \over k-2} - w \tilde{m} -{kw^2 \over 4}
+ \tilde{N} + h -1 \right) |\tilde{j}, \tilde{m}, \tilde{N}, h\rangle = 0 \cr
L_n |\tilde{j}, \tilde{m}, \tilde{N}, h\rangle 
& = (\tilde{L}^{AdS}_n + L^{\cal M}_n -w \tilde{J}^3_n)  
|\tilde{j}, \tilde{m}, \tilde{N}, h\rangle =0 \;, \ \ \ \ n \ge 1
}}  
and obtain the same expression for the energy that was found in 
$AdS_3 \times {\cal M}$, \shorte .  The discussion for the continuous
states is similar and once again we conclude that the energy is given by
\longe .

Normally, twisting the currents as in \nsector\ gets rid of the zero
mode and the corresponding  total charge 
$Q^{\pm}={1 \over 2\pi} \int J^{\pm}d\sigma$ vanishes.  
This results in breaking of the gauge symmetry \refs{\vafwit, \govin}.  
What we have found here, in the case of $AdS_3$, is that such twists are
nothing but fractional spectral flows.  One might worry that there is still 
a distinction between those states built with integrally moded $J^{\pm}$ and 
those states built with fractionally moded $J^{\pm}$, in that the latter are 
expected to have a different ground state energy.  However, in the next
section we will show from the partition function calculation that this
does not happen.  As such, by taking  
$\hat{{\cal D}}^{+}_{\tilde{j}}$, $\hat{{\cal C}}^{\a}_{1/2 +is}$ and 
their images under fractional spectral flows, we automatically include 
the twisted states.  Of course, the integer-valued spectral flows are
still allowed and all the flowed sectors are treated in equal footing.
In particular, the form of the Virasoro constraints remains the same and
so does the expression for the energy and angular momentum.
It is tempting to think that even in the case of $AdS_3$, spectral flow 
arises as a kind of twisting of some underlying theory, possibly with
$\phi$ noncompact.  But one probably needs a better understanding of the 
$SL(2,R)/U(1)$ parafermion theory \refs{\lyk} in order to pursue this idea. 

\subsec{Invariant subspace}

Having constructed the twisted sectors, only the states that are invariant
under the identification $\phi \sim \phi + 2 \pi/N$ 
are to be retained in the spectrum.  
There is a simple way to see  what one should expect.  
If one considers the wave equation for a scalar field
in the background \metric , the solution may be expressed as 
$\Psi = \sum R (r, \omega, m) e^{-i \omega t+ im \phi}$.  
Then single-valuedness of the wave function implies 
$m = N \times {\rm integer}$.
The effect of the projection, then, is to restrict the angular momentum
to be a multiple of $N$.  

It is straightforward to see how this condition comes about.
For all the sectors that we have, we are to project on to the states invariant
under the operator
\eqn\twi{
e^{-2 \pi i (J^3_0-\bar{J^3_0})/N} \;.
}
Therefore, the states that remain satisfy the condition 
$\ell = J^3_0-\bar{J^3_0} = N \times {\rm integer}$, 
for both the discrete and continuous representations.

\subsec{Thermal partition function}

As in the case of $AdS_3$, we can check that the spectrum derived above
agrees with what one gets by evaluating the finite temperature partition 
function.  The calculation in the case of conical spaces is a minor 
extension of the calculation in \refs{\mos} for $AdS_3$ and our
focus will only be on the effects due to the conical singularity.  
The reader is referred to that paper for greater detail.

We first transform to the coordinates that are well suited for carrying out
the path integral.  They are given by \refs{\gawed} 
\eqn\prtcrd{\eqalign{
v       & =  \sinh \rho \ e^{i \phi}  \cr
\bar{v} & =  \sinh \rho \ e^{-i \phi}  \cr         
\theta    & =  t - \log \cosh \rho \;.
}}
Under the identification $\phi \sim \phi + 2 \pi /N$, the fields
are identified as $v \sim v e^{2 \pi i /N}$ and 
$\bar{v} \sim \bar{v} e^{-2 \pi i /N}$.  We take the worldsheet to be 
a torus with modular paramter $\tau$.  Then the boundary conditions are
\eqn\vbc{
v(z+2 \pi) = v(z) e^{2\pi i a/N} \;, \ \ \ \ 
v(z+2\pi \tau) = v(z)e^{2\pi i b/N} \;.
}
We will denote by ${\cal Z}_{ab}$ the path integral 
$\int e^{-S} {\cal D}\theta {\cal D}v {\cal D}\bar{v}$ with the above 
boundary conditions.  We remind the reader that these boundary
conditions are in addition to those introduced by identifying the Euclidean
time $t \sim t + \beta$.

Let us first calculate ${\cal Z}_{a0}$.  We can implement the right
boundary condition by setting
\eqn\vthc{
v(z) = \tilde{v} \exp 
\left(-{a \over 2N \tau_2}(z\bar{\tau} - \bar{z}\tau)\right) \;,
}
with $\tilde{v}$ periodic.  Then $\bar{U}_{n,m}$, as defined in eqn.\ (23)
of \refs{\mos}, picks up an additional term, 
$\bar{U}_{n,m} \to \bar{U}_{n,m} + a\bar{\tau}/N$.  With this change, we
can repeat the calculation that was done in \refs{\mos}, and obtain the 
partition function as written in eqn.\ (27) of that paper:
\eqn\cpart{\eqalign{
{\cal Z} = \ & {\beta \over 8 \pi}
\left( {k-2\over\tau_2} \right)^{\frac12} \ \times \cr
 & \sum_{n,m} 
{ e^{- k \beta^2 | m - n \tau|^2/ 4\pi\tau_2
+ 2\pi({\rm Im}  U_{n,m})^2 /\tau_2}  \over 
|\sin(\pi U_{n,m})|^2 |\prod_{r=1}^\infty 
(1-e^{2\pi i r\tau})(1-e^{2\pi i r\tau+ 2\pi iU_{n,m}})
(1-e^{2\pi i r \tau-2\pi iU_{n,m}})|^2 } \;.
}}
Similarly, for 
${\cal Z}_{0b}$  all we need to do is twist along the other direction 
of the torus, to obtain $\bar{U}_{n,m} \to \bar{U}_{n,m} + b/N$ and once gain
the partition function takes the same form.  
In this way we obtain for the partition function of thermal $AdS_3/Z_N$,
\eqn\tpart{
{\cal Z} = {1 \over N} \sum_{a, b} {\cal Z}_{ab} \;.
}

To obtain the free energy of strings on $AdS_3/Z_N \times {\cal M}$, 
we multiply
\tpart\ by the partition function of the CFT on ${\cal M}$ and the 
reparametrization ghosts, and integrate $\tau$ over the fundmental domain:
\eqn\free{
\int_{F_0}{\cal Z}_{AdS_3/Z_N}{\cal Z}_{\cal M} {\cal Z}_{bc}
=-\beta F = -\sum_{physical} \log (1-e^{-\beta E}) \;.
}
However, as shown in \refs{\polchin, \mcclain}, one can obtain an 
equivalent expression where the integration is over a larger domain while
restricting the sum in \cpart\ to $n=0$.  
Then one can follow exactly
the same steps as in \refs{\mos} to reproduce the spectrum.  
We will explain some
of the new features that arise in the course of this computation.

As usual the sum over $a$ represents the twisted sectors and the
sum over $b$ serves as a projection down to the invariant states. 
Consider ${\cal Z}_{a0}$ and its expansion along the lines of \refs{\mos}.
From $\exp \{2\pi({\rm Im} U_{0,1})^2/\tau_2\}$ and
$|\sin (\pi U_{0,1})|^{-2}$ we obtain the additional factor
\eqn\shge{
\exp\left[2\pi \tau_2 \left({a^2 \over N^2}-{a\over N} \right)\right]\;.
}
A conformal field theory of 2 bosons with periodicity $\theta$ has
ground state energy
\eqn\tgse{
(q\bar{q})^{-{1\over 2}(\theta^2 -\theta) -{1\over 12}}\;,
}
so we have reproduced what might have been the expected shift in the 
ground state energy.  However, this is not the end of story.  
The oscillator terms are changed to
\eqn\cosc{
\left| \prod_{n=1}^{\infty} (1-e^{{\beta} + 2 \pi i\tau (n-a/N)})
(1-e^{-{\beta} + 2 \pi i \tau (n+a/N)}) \right|^{-2}\;,
}
which has poles when $\tau_2 = {\beta \over 2\pi(n-a/N)}$. In \refs{\mos} 
it was shown that the location of the poles correspond to spectral flow
parameters.  So we see that $w$ is given by $w =n-a/N$ with $n$ being positive
integers. It will be explained shortly that $w=-a/N$ arises from $\tau_2$
above the first pole at ${\beta \over 2\pi(1-a/N)}$. 
The shift in the location of the poles also causes the expansion of 
\cosc\ to be slightly different from the $AdS_3$ case.  
One finds the terms (compare to eqn.\ (47) of \refs{\mos})
\eqn\ecosc{
\dots \exp \left[2 \pi \tau_2 \left(w(w+1) 
- {a^2 \over N^2} + {a \over N} \right) \right] \dots
}
The extra terms on the right serve to cancel the shift in ground state
energy, \shge , and we are left with the correct expression for the energy.  
Note that this cancellation is in agreement with what we found in the previous
section. What appears to be twisting is actually a fractional spectral flow.  

To see that summing over $b$ corresponds to a projection down to the 
invariant states, take ${\cal Z}_{ab}$ and its expansion.  The only 
additional change is the appearance of a new term
\eqn\phab{
\exp \left[ -{2\pi i b(q-\bar{q}) \over N} \right]
}
in every state.  Hence, ${1 \over N}\sum_b {\cal Z}_{ab}$
only includes the states with the correct condition on angular momentum.
This shows that from \tpart\ we obtain the spectrum that agrees with
what was found in the algebraic analysis.

\subsec{Bound on $\tilde{j}$}

In expanding the partition function, the presence of poles in 
the oscillator terms meant that the range of $\tau_2$ was broken up into 
\eqn\taur{
{\beta \over 2 \pi(w+1)} < \tau_2 < {\beta \over 2 \pi w} \;,
}
and a different expansion was used in each interval.  
This gave rise to the states with spectral flow by amount $w$.   
In the case of $AdS_3$, this included the sector with $w=0$.  But
now that $w$ is no longer limited to be an integer, we need to re-examine 
the special case 
\eqn\tauinf{
{\beta \over 2\pi(1-a/N)} < \tau_2 < \infty \;.  
}
In this range,
the energy is found to be 
\eqn\efp{
E = 1 + q + \bar{q} - {2a \over N} + \sqrt{1+4(k-2) 
\left(N_w +h -1 - {1 \over 2} 
\left({a^2 \over N^2}-{a \over N}\right) \right)} \;.
}
So we see that these states are in the sector flowed by $w = -a/N$.  
Thus, the allowed values of spectral flow are $w=n-a/N$, including $n=0$.  We
expect that these states will have a different range of $\tilde{j}$, because
the integral over $\tau_2$ is broken up in a different way from all the other
states.\foot{That is to say, in the variable $1/\tau_2$ these states
occupy a strip of length less than $2\pi /\beta$.} 
Repeating the saddle point calculation as was done in \refs{\mos},
$\tilde{j}$ is seen to satisfy
\eqn\nbj{
{(k-2){a \over N} + 1 \over 2}< \tilde{j} < {k-1 \over 2} \;.
}
On the algebraic side, this change in the lower bound can be seen from solving
the physical state condition \physt
\eqn\psj{
\tilde{j}=\frac12 -{k-2 \over 2} w + \sqrt{\dots} \;,
}
with $w=-a/N$.  The semi-classical limit (large $k, h$) of this bound 
translates into
\eqn\scbj{
0 < \sqrt{{4h \over k}} < 1 - {a \over N} \;,
}
which is consistent with the analysis of \refs{\mo}, extended to negative
values of $w$ (see section 3 of that paper).  In $AdS_3$, states with
negative $w$ automatically had negative energy, but now we find that in the
quotient space it is possible for states with negative fractional spectral 
flow to have positive energy.

\newsec{Summary and discussion}

We have formulated a description of strings moving on $AdS_3$ but with 
an opening angle of $2 \pi/N$ for $\phi$.  
The twisted states arising from the
orbifold construction found a natural description as states with 
fractional spectral flow.
Specifically, we have shown that the $n$th twisted sector is obtained by
taking spectral flow with $w=n/N$.  Rather than thinking of the original 
states with integral $w$ as being ``untwisted'' and fractional $w$ as being
``twisted'', we have proposed that there is only one untwisted sector, 
namely those with $w=0$,  and
all the spectral flowed sectors should be thought of as being twisted.

We have also computed the thermal partition function on this background and
extracted the spectrum that agrees with the results of the algebraic 
description.  Despite the fact that there are states constructed by acting
with fractionally-moded generators, it was shown that this does not cause a 
change in the ground state energy.   
  
The fact that twisted states may be obtained by spectral flow means that
we are also able to write down the corresponding vertex operators, by
bosonizing the $J^3$ current \refs{\gep, \mo}.  Thus, unlike what usually
happens in orbifolds we have explicit formulas for the twisted state vertex 
operators.\foot{See \refs{\mart} for a detailed discussion.}  Using these
vertex operators we can compute the long string scattering amplitude on
$AdS_3/Z_N$, following \refs{\mo}.
 
One might wonder whether we can extend our analysis to the case with 
rational values of the opening angle.  
Indeed, it is fairly simple to generalize
the algebraic construction given here, by first going to the covering space
in which $\phi$ has period $2\pi P$ and taking a $Z_Q$ orbifold.  The
resulting space would have an opening angle $2 \pi P/Q$.  However,
it is not clear whether one can calculate the partition function with this
geometry, and that prevents us from concluding at present that such 
descriptions are possible.

\

\centerline{\bf Acknowledgements} I am grateful to J. Maldacena for 
helpful discussions and comments on the manuscript.
This work was supported in part by DOE grant DE-FG02-91ER40654.
\listrefs

\end